# Structural diversity of molecular nitrogen on approach to polymeric states

Alexander F. Goncharov[1], Iskander G. Batyrev[2], Elena Bykova[3], Lukas Brüning,[4] Huawei Chen[1,5], Mohammad F. Mahmood [5], Andew Steele[1], Nico Giordano,[6] Timofey Fedotenko[6], Maxim Bykov[4,7]

[1] Earth and Planets Laboratory, Carnegie Institution for Science, Washington, DC 20015, USA
[2] U.S. Army Research Laboratory, FCDD-RLW-WA, Aberdeen Proving Ground, Maryland 21005, USA
[3] Goethe-Universität Frankfurt am Main, Facheinheit Mineralogie, 60438 Frankfurt am Main, Germany
[4] Institute of Inorganic Chemistry, University of Cologne, Greinstrasse 6, 50939 Cologne, Germany
[5] Department of Mathematics, Howard University, Washington D.C. 20059, USA
[6] Deutsches Elektronen-Synchrotron DESY, Notkestr. 85, 22607 Hamburg
[7] Institute of Inorganic and Analytical Chemistry, Johann Wolfgang Goethe Universität Frankfurt, Max-von-Laue-Straße 7, D-60438 Frankfurt am Main, Germany

**Nitrogen represents an archetypal example of material exhibiting a pressure driven transformation from molecular to polymeric state. Detailed investigations of such transformations are challenging because of a large kinetic barrier between molecular and polymeric structures, making the transformation largely dependent on kinetic stimuli. In the case of nitrogen, additional complications occur due to the rich polymorphism in the vicinity of the transition. Here, we report the observation of both molecular (θ) and polymeric (BP) phases, crystallized upon temperature quenching of fluid nitrogen to room temperature at 97-114 GPa. Synchrotron single-crystal X-ray diffraction, Raman spectroscopy, and first-principles theoretical calculations have been used for diagnostics of the phases and determination of their structure and stability. The structure of θ–$N_2$ is determined as tetragonal, space group $P4_22_12$—one of the phases previously predicted theoretically above 9.5 GPa. Molecular θ–$N_2$ is the most stable among molecular phases bordering the stability field of polymeric phases, partially settling a previously noted discrepancy between theory and experiment concerning the thermodynamic stability limit of molecular phases.**

Pressure-induced polymerization is a universal phenomenon in simple molecular systems. It occurs due to the destabilization of strong intramolecular bonds, where at high compressions a large electron density becomes energetically less favorable than that of the network lower order bonding. Well-known examples are $N_2$ [1, 2], $CO_2$ [3], $H_2O$ [4], $I_2$ [5], etc. The case of nitrogen is of fundamental interest because of the simplicity of the system, and a very high bonding energy of the intramolecular triple bond, which makes polymeric all-nitrogen compounds extremely energetic (see Ref. [6] and Refs. therein). Thus, knowledge of the structure and understanding of the stability limits of the phases bordering the molecular-polymeric transition is critical. This information remains at least partially controversial because of a large potential barrier between molecular and polymeric phases, which results in large kinetic effects such as large hysteresis limits depending on temperature and direction of the transformation. Thus, challenging experiments need to be performed in a wide pressure-temperature (P-T) range (50 to 150 GPa and 10 to 7000 K). In addition, molecular and polymeric phases are characterized by extremely rich polymorphism, which includes the realization of many metastable phases (see Ref. [7] and Refs. therein).

Density functional theory calculations predict the transformation from molecular to nitrogen allotropes above 50 GPa [1, 8-12]. However, experiments show that the transformation to a polymeric amorphous phase η only occurs at much higher pressure (>120 GPa) at temperatures <2000 K [13-15]. Higher temperatures are needed to crystallize polymeric nitrogen in cubic gauche (cg) and black phosphorus (bp) structures, but the synthesis can be performed at lower pressures [2, 12, 16, 17]. Furthermore, fluid molecular nitrogen solidifies into cg-N at 116 GPa and 2080 K and at 120 GPa and 2440 K [18] upon isothermal compression. No polymeric liquid is reported below 2440 K, and no solid polymeric phase forms below 116 GPa, e.g., by crystallization of fluid molecular nitrogen across its melting line. Polymeric nitrogen likely forms at higher temperatures above 125 GPa, evidenced by an appearance of a Drude like reflectance edge [19].

The discrepancy in polymerization pressure between experiment and theory has previously been addressed in a number of ways. Proper inclusion of van-der-Waals interactions allows to widen the thermodynamic domain of stability of molecular phases [20]. Also, the entropic effects tend to stabilize the molecular phases [11, 21], thus shifting the transformation pressure to higher values; however, high temperatures are needed to overcome the potential barriers. Another factor, which could lead to the large discrepancy in the transformation pressure, is the possible existence of another more stable molecular phase near the border with the stability domain of polymeric phases [8]. Reliable structural data have been reported for ε-nitrogen— high-pressure phase which is stable in a wide P-T range from 17 GPa at 300 K to 115 GPa and 2000 K (Fig. 1), where it is expected to be limited by the equilibrium transition line with the domain of stability of polymeric phases. ε-nitrogen transforms to the ζ-phase at 65 GPa at room temperature. ζ- nitrogen is inferred to have a slightly distorted structure of the rhombohedral ε -$N_2$ but its structure has not been determined definitively[22]. ζ-nitrogen transforms to ε-nitrogen at high temperature along the phase line with a positive slope (Fig. 1). Another molecular high-pressure phase (κ), which was hypothesized to represent another lowering of the unit cell symmetry of ε -$N_2$, is reported above 110 GPa [22]. It is interesting that heating ε, ζ, and κ phases above 56 GPa across or close to the melting line results in the formation of other molecular phases: ι at 56-69 GPa [9, 23] and ζ′ phase above 75 GPa [24, 25]. The latter ζ′ phase is also a product of the pressure release transformation from the amorphous polymeric phase η at high temperature [24]. It should be noted that according to theoretical calculations ε -$N_2$ is metastable. Instead, random structure searches revealed several other energetically more favorable molecular phases [8]; some of these predictions have been confirmed, for example for λ-nitrogen [26]. Yet another molecular phase, θ-nitrogen, has been reported to occur above 90 GPa at high temperatures (700-925 K) [23]; which has very similar Raman spectra to λ-nitrogen [26]. Its formation is irregular and usually require P-T manipulations, e.g., isothermal compression at high T [25]. The structure of this phase has not been determined but is assumed to be orthorhombic based on preliminary powder XRD data [23].

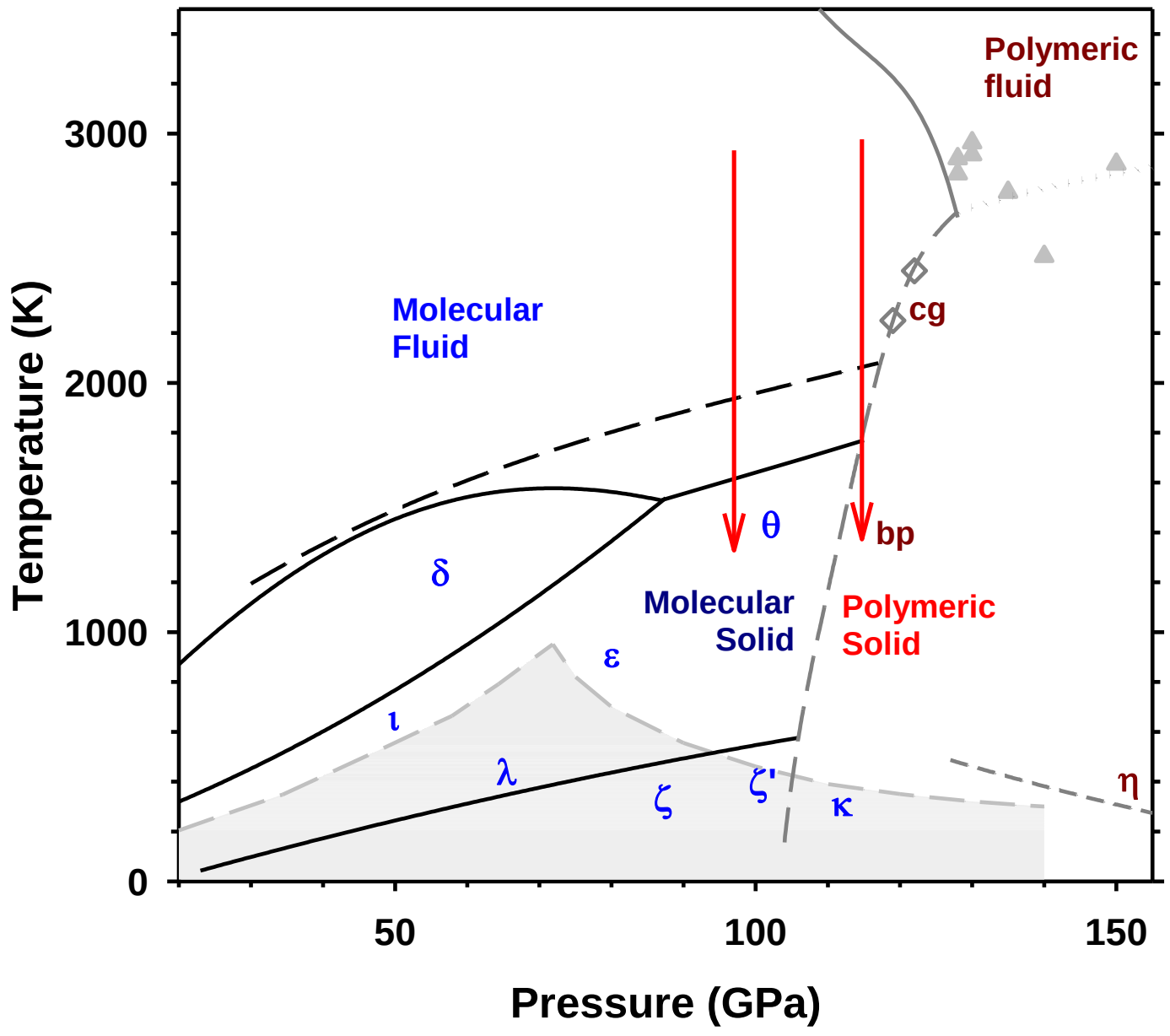

**Figure 1.** Phase diagram of nitrogen at extreme thermobaric conditions. Solid black lines are the phase and melting lines of molecular phases from Refs. [26, 27]. The melting line from Ref. [18] is shown for comparison as a dashed black line. Short gray dashed line is a kinetic line along which a molecular high-pressure phase κ transforms to an amorphous phase η [13, 15, 24]. Long dashed gray line is the proposed in Refs. [18, 27] phase line between molecular and polymeric solids. Dotted gray line is a hypothetical melting line of polymeric phases, where metallization occurs according to Ref. [19]. The onset to the conductive molecular dissociative state is shown by a solid gray lime [19]. The solidification pathways of this work, which resulted in crystallization of molecular θ-nitrogen and polymeric bp-nitrogen are shown by vertical red arrows.

Here we report the structural determination of θ-nitrogen, which was synthesized by quenching of laser heated nitrogen above the melting line at 97 GPa. The structure of this phase is tetragonal $P4_12_12$ with four molecules in the unit cell. Theoretical structure search found this molecular phase to be the most stable above 9.5 GPa [8], which was confirmed by theoretical calculations of in this work. Density functional theory calculations with norm-conserving pseudopotentials show that this phase transforms to cg-N at 70 GPa, which is 5 GPa higher than the calculated transformation pressure from ε-nitrogen to cg-N. The experiments show the transformation of $P4_12_12$ $N_2$ to bp-N at 117 GPa compared to the theoretically predicted 81 GPa. This difference is likely due to an increased stability of molecular nitrogen at high temperatures, which are needed to overcome the kinetic barrier.

The experimental procedure included concomitant single crystal X-ray diffraction (SCXRD) and Raman spectroscopy measurements at 97–114 GPa at the Extreme Conditions Beamline (ECB) at Petra III (DESY, Hamburg) and the following up Raman mapping at our Carnegie Science facilities. A small flake of metallic Co was positioned in the DAC cavity to absorb heat from an infrared laser. The quenched samples were mapped at room temperature using synchrotron powder XRD to identify points of interest with good single-crystal character which were then examined by SCXRD.

For single-crystal XRD measurements at ECB, we used monochromatic X- ray radiation with λ = 0.2908 Å focused down to 2×2 $\mu m^2$ by a Kirkpatrick–Baez mirror system and diffraction patterns were collected on a Perkin Elmer XRD1621 flat panel detector. For the single-crystal XRD measurements, samples were rotated around a vertical ω-axis in a range of ±30° with an angular step Δω = 0.5° and an exposure time of 1-2 s/frame. For analysis of the single-crystal diffraction data we used the CrysAlisPro software package [28], which facilitates the SCXRD analysis of multigrain samples. To calibrate an instrumental model in the CrysAlisPro software, i.e., the sample-to-detector distance, detector's origin, offsets of goniometer angles, and rotation of both X-ray beam and the detector around the instrument axis, we used a single crystal of orthoenstatite ($(Mg_{1.93}Fe_{0.06})(Si_{1.93}, Al_{0.06})O_6$, *Pbca* space group, *a* = 8.8117(2), *b* = 5.1832(1), and *c* = 18.2391(3) Å). The structure was solved with the ShelXT structure solution program and refined with the Olex2 program[29, 30]. The Cambridge Structural Database [31] contains the supplementary crystallographic data for this work (CSD-2299301). These data can be obtained free of charge from FIZ Karlsruhe [32].

First-principles calculations were performed within the framework of density functional theory (DFT), implemented in Vienna Ab initio Simulation Package (VASP) code [33]. We implemented Perdew-Burke-Ernzerhof (PBE) [34] functional for the exchange-correlation interactions of electrons. Plane-wave basis with a cut-off energy of 500 eV and sample K-point of 2π*0.025 $Å^{-1}$, were used to calculate total energy. We relaxed the crystal structure until all the stress forces of atoms were smaller than 0.01 eV/ Å. We used the forces from the VASP code to calculate the phonons. First-principles theoretical calculations have been performed in $P4_12_12$ (θ), $R\bar{3}c$ (ε),

*Cmce* (bp), and $I2_13$ (cg) phases at selected pressures (60 and 112 GPa), where these structures were optimized using norm-conserving pseudopotentials and GGA-PBE functional. Monkhorst-Pack grid size for k-points sampling of the Brillouin Zone (BZ) is 5x6x5 for all structures [35]. The phonon dispersion and phonon frequency calculations were performed using a finite displacement method implemented in the CASTEP code[36]. The Raman spectra were calculated using the formalism presented in Ref. [37]. The electronic bandgap calculations have been performed within GGA/PBE approximations.

Laser heating to 2700 K at 97 GPa results in formation of another phase other than the initial ζ phase (Fig. 2), which we identified as θ-nitrogen [23] based on XRD and Raman measurements as described below. Visual observations show the presence of a strip at the edge of Co heat absorber after heating, suggesting the presence of material with a different refractive index (Fig. S3, Supplements[38]. Single-crystal XRD pattern (Fig. 2(a), and Table SI, Supplements[38]) can be indexed in the $P4_12_12$ space group (# 92, $a$ = 2.6943(5), $c$ = 7.52(3)), and the structure can be solved and refined as molecular nitrogen with four $N_2$ molecules in the unit cell with an N-N intramolecular distance of 1.09(1) Å. A single crystallographically-independent nitrogen atom occupies a general position 8*b* (0.2907(9), 0.5578(10), 0.4740(14)).

The structure can be viewed as layered, where the layers are formed by canted molecules connected by stronger intermolecular forces compared to the interlayer ones (Fig. 2(b) and Fig. S1, Supplements[38]). The unit cell consists of four layers stacked in a sequence ABCD; the structure can be described as a distorted *fcc* lattice [8]. It is remarkable that another molecular phase, λ-$N_2$ [8, 26], which has a monoclinic $P2_1/c$ structure, has similar structural motif consisting of essentially the same layers as in θ-$N_2$ (Fig. S1, Supplements[38]). However, the layer stacking sequence in λ-$N_2$ differs in that it only has two layers (AB) in the unit cell. The volumes occupied by the molecule would be expected to be similar for these two molecular structures, which can be considered as polytypes. This is indeed the case as the experimental volume vs pressure curves for these phases can be described by the same equation of state (Fig. S2, Supplements[38]). The structural similarities between these polytypes give rise to similar Raman spectra of the two modifications.

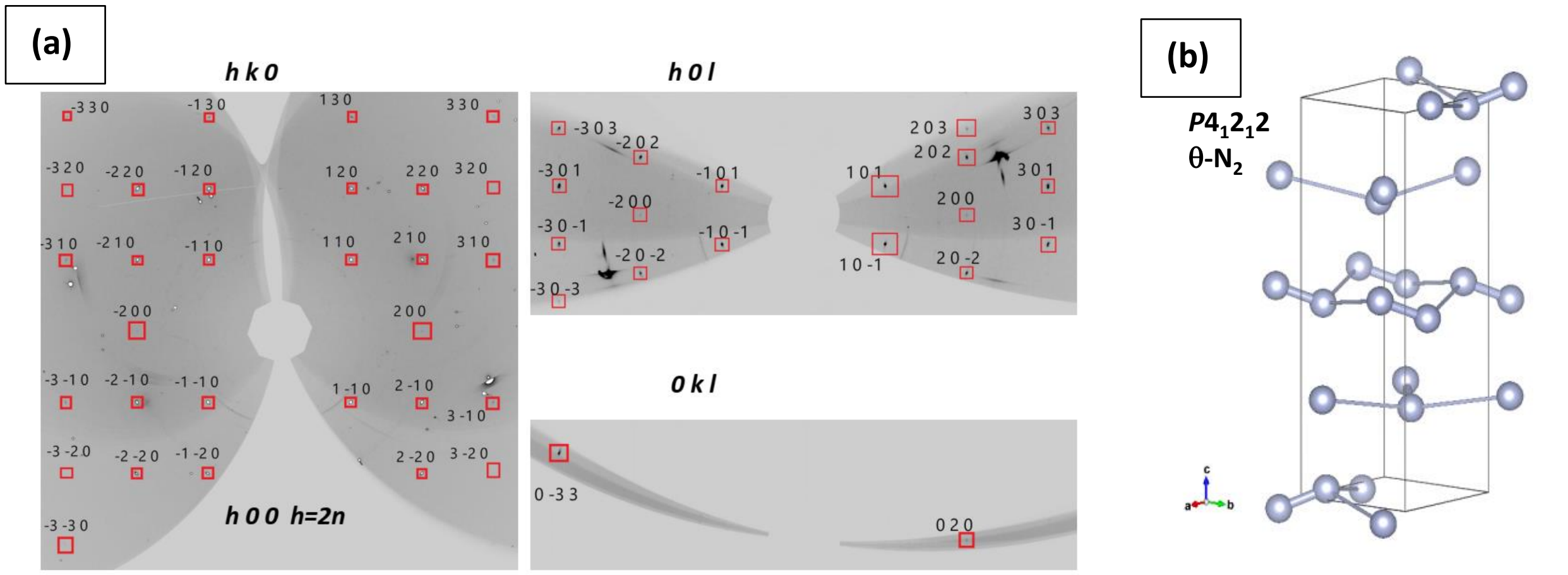

**Figure 2. (a)** Reconstructed reciprocal lattice planes of θ-nitrogen at 97 GPa. Observed diffraction spots from sample have been indexed and used to determine the structure of $P4_12_12$ $N_2$ (see details in Table SI, Supplements[38]). The X-ray extinctions rule for this phase is depicted; (b) Crystal structure of θ-$N_2$. Thick lines represent intramolecular N-N and thin lines represent intermolecular bonds.

The Raman spectra of the new phase corresponds well to that of θ-nitrogen [23, 25], characterized by low-frequency translational and rotational modes and high-frequency intramolecular modes (Fig. 3). Group-theoretical analysis (Table S2, Supplements[38]) predicts rich Raman spectra, which consist of 3 high-frequency intramolecular modes (N-N stretching vibrations), 6 translational, and 5 librational Raman modes (E-modes are doubly degenerate). Our Raman experiments demonstrate good agreement with this prediction (Fig. 3). The theoretically calculated Raman spectrum also shows good qualitative agreement in the number of modes and composition of the measured spectrum (Fig. 3). As expected, the Raman spectrum of θ-nitrogen is very similar to λ-nitrogen [26]; the difference is due to the larger unit cell of θ-nitrogen, which results in a larger number of modes to be Raman active (Table S2, Supplements[38]). These include 1 intramolecular mode, 6 low-frequency translational modes, and 3 librational modes. Our experiments indeed show that there is a correspondence between the strongest Raman bands in both phases, while θ-$N_2$ reveals one extra $N_2$ vibron mode and 6 extra translational and librational modes in the low-frequency range (Fig. S4, Supplements[38]). This is in good agreement with the predictions since some of the additional modes may be weak or interfere with stronger modes. The strongest Raman peaks of θ-$N_2$ and the Raman peaks of λ-$N_2$ demonstrate almost identical pressure dependencies of the frequencies (Fig. S4, Supplements[38]).

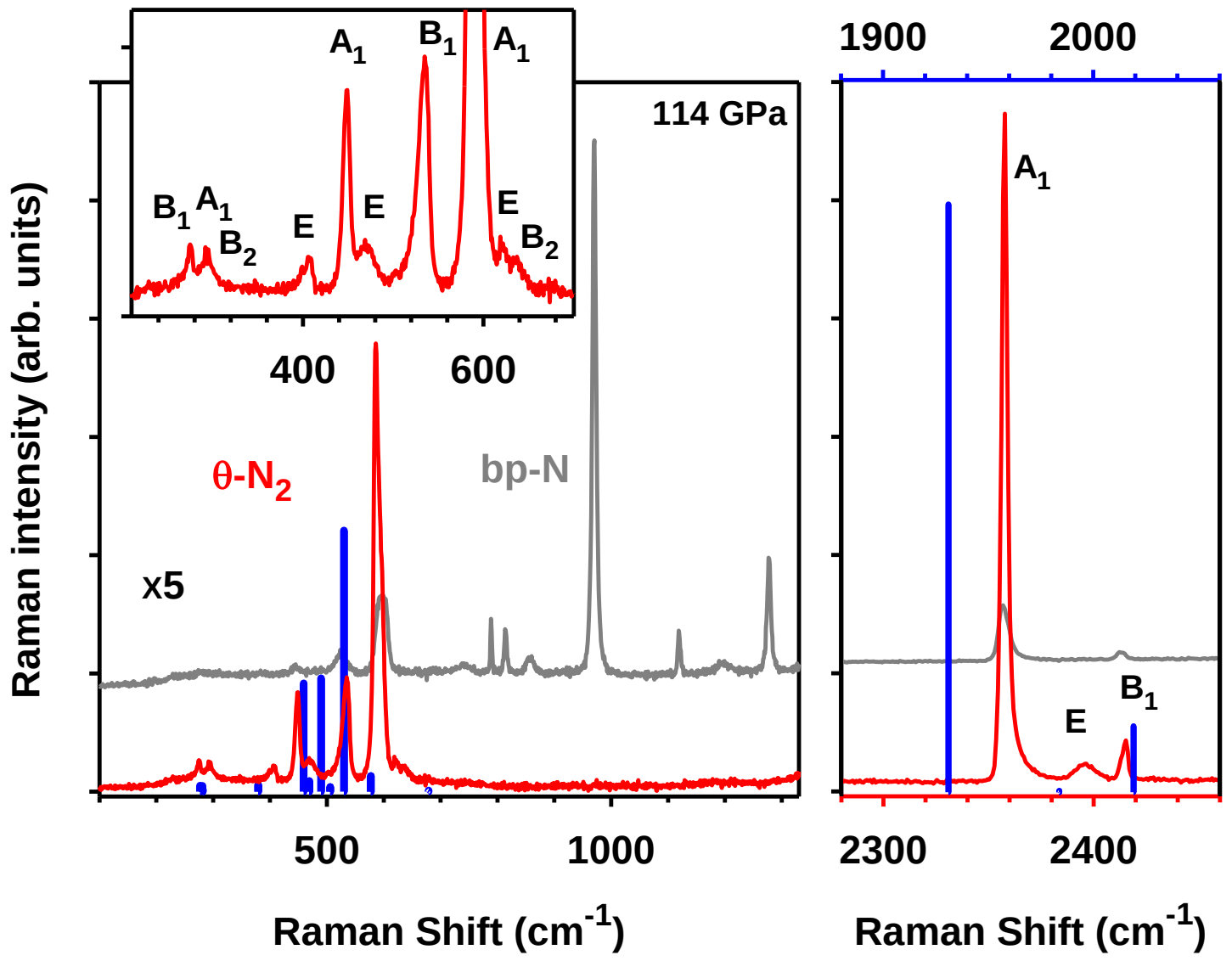

**Figure 3.** Raman spectra of θ-$N_2$ and bp-N at 114 GPa. The top trace, which nominally corresponds to bp-N, contains peaks of θ-$N_2$ due to the proximity of this phase in the sample cavity. The theoretically calculated Raman spectrum at 112 GPa is depicted by blue solid vertical bars. The inset shows an expanded view of the experimental low-frequency Raman spectrum. The Raman peaks are assigned according to group-theory (Table S2, Supplements[38]) and theoretical calculations.

The sample was reheated at each of the three successive higher-pressure points. At the highest pressure of this run (114 GPa), polymeric bp-N [12, 17] was detected in powder XRD patterns (as single-crystal like reflections) (Fig. S5 of Supplements[38]) and in Raman spectra (Fig. 3) at one of the positions, which had been occupied by θ-nitrogen before heating. The lattice parameters, the number of Raman peak and their spectral positions of bp-N (Fig. 3, Figs. S2, S4, Supplements[38]) agree well with the previous observations [12, 17].

Our theoretical calculations demonstrate that the experimentally determined structure of θ-$N_2$ relaxes to the phase with the same symmetry and essentially identical fractional atomic coordinates. This phase is mechanically and dynamically stable at 112 GPa (Fig. S6, Table S3,

Supplements[38]); it is semiconducting with a band gap of 2.23 eV (Fig. S7, Supplements[38]). We computed the enthalpy of θ-$N_2$, ε-$N_2$, bp-N, and cg-N at 60 and 112 GPa (Fig. 4). In this pressure range θ-$N_2$ is more stable compared to ε-$N_2$, so the equilibrium transition to the most stable polymeric cg-N phase shift up to 70 GPa vs 65 GPa. Our calculations show that another polymeric phase bp-N is slightly less stable than cg-N; thus the θ-$N_2$ to bp-N transition is expected to occur at even higher pressures, at 80.7 GPa. These calculations correspond to 0 K, and the results should be corrected for the entropy terms, so the presented results should be considered as an approximate guide and must be temperature corrected to be quantitatively compared with experiments. This correction can be approximated by accounting for the change in the chemical potential of nitrogen at high pressure-temperature conditions [11]. Using the approximation of "the Moderate Extrapolation" of Ref. [11] and assuming that the synthesis of θ-$N_2$ occurred near the melting line near 1750 K, this approximation results in the shift of the transition pressure to cg-N to 104 GPa, (112 GPa to bp-N) which is close to the experimental conditions.

The results of our enthalpy vs P calculations (Fig. 4) qualitatively agree with the previous calculations [8, 9], which found that $P4_12_12$ $N_2$ is among the most stable molecular phases above 10 GPa. These calculations also show that λ-$N_2$ is almost equally stable as θ-$N_2$, while ε-$N_2$ is distinctly less stable. This result seems to disagree with numerous experiments, which show that ε-$N_2$ and related higher-pressure phases (ζ and κ) are stable or metastable at room temperature upon compression at room temperature to the limit of stability of the molecular phases. However, given the wide region of P-T space where both λ-$N_2$ and θ-$N_2$ polytypes can be observed especially at low T [23, 26], the stability of ε-$N_2$ and derived phases must be driven by the entropy and pressure-volume (PV) terms the Gibbs free energy (Fig. S2, Supplements[38]) below 50 GPa, while at high P, these phases are likely metastable, confirmed by their amorphization on the pressure increase at low temperatures and crystallization to ι-$N_2$ and θ-$N_2$ upon quasi-isobaric heating or crystallization from the melt.

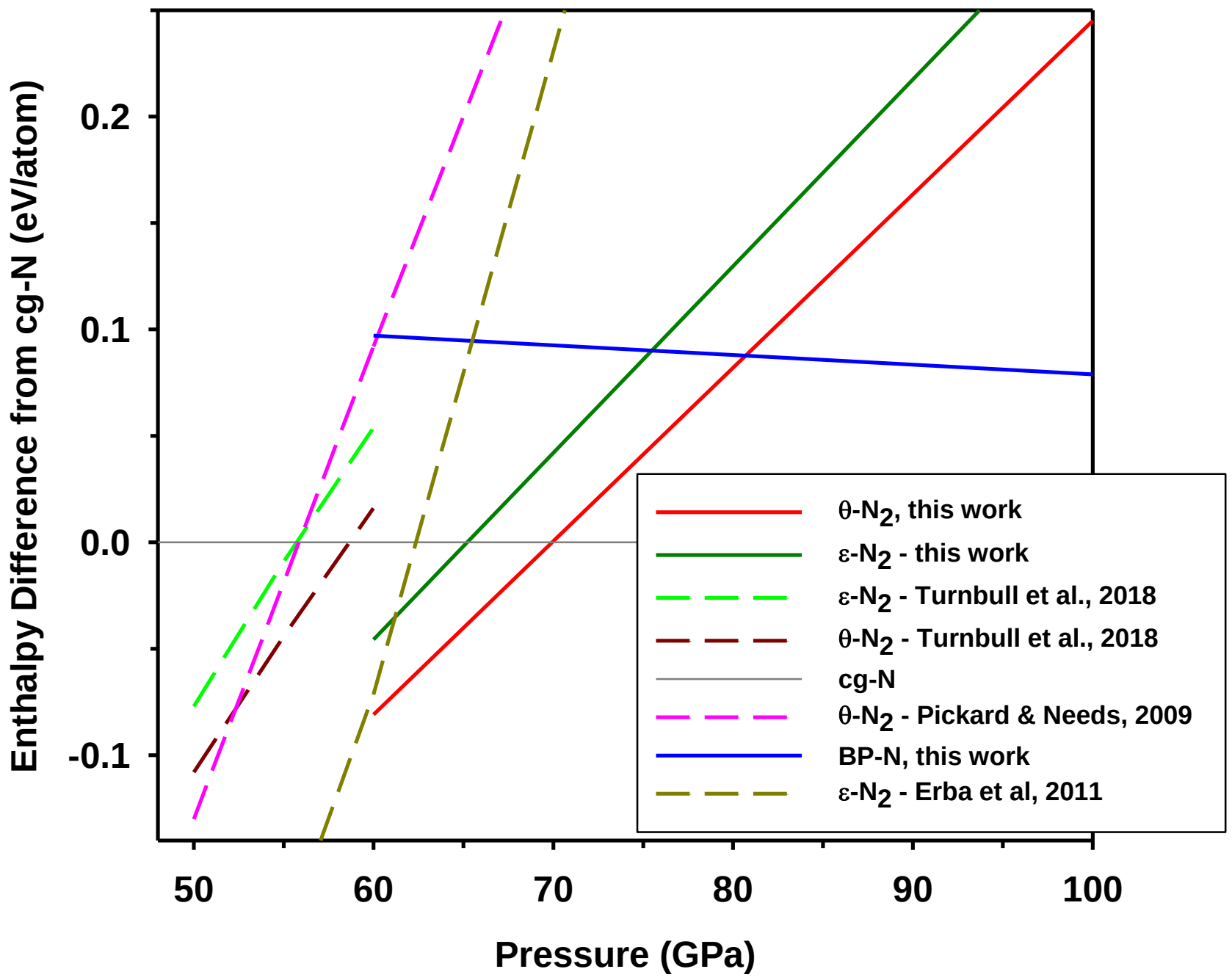

**Figure 4.** Theoretically computed enthalpies of molecular and polymeric nitrogen phases vs pressure determined here from the first principles plotted with respect to results for cg-N. Solid lines are the results computed here while dashed lines are the results of previous computation [8, 9, 20]. The results of this work show the transitions to polymeric phases at higher pressures because of using norm-conserving pseudopotentials.

Our experiments show the previously unobserved transformation of θ-$N_2$ to bp-N at 114 GPa. Even though we cannot completely rule out recrystallization from the melt, this observation sets an important transformation pressure-temperature condition for the molecular to polymeric phase transition near 114 GPa and 1750 K in fair agreement with previous observations [18] and conjectures [27]. It is interesting that θ-$N_2$ transforms to bp-N polymeric phase instead of believed to be thermodynamically stable cg-N. We propose that this is because of the kinetic reason as the structures of θ-$N_2$ and bp-N phases bear similarities, both consisting of puckered layers (Fig. S1, Supplements[38]).

Overall, our combined experimental and theoretical investigations demonstrate the previously missed transformation pathway between molecular and polymeric states of nitrogen. The P-T conditions of the transformation settles an equilibrium transition line above 110 GPa, well above theoretically calculated at 0 GPa (55-70 GPa). The difference can be understood in terms of the previously proposed change in the chemical potential of nitrogen at high P-T conditions.

A.F.G., H.C., and M.F.M acknowledge the support of the National Science Foundation (NSF) Grant No. DMR-2200670. M.B. acknowledges the support of Deutsche Forschungsgemeinschaft (DFG Emmy-Noether Program project BY112/2-1). E.B. acknowledges the support of Deutsche Forschungsgemeinschaft (DFG Emmy-Noether project No. BY 101/2-1). We acknowledge DESY (Hamburg, Germany), a member of the Helmholtz Association HGF, for the provision of experimental facilities. Parts of this research were carried out at PETRA III (beamline P02.2). Beamtime was allocated for proposals I-20221160 and I-20230233.

Supplemental Information

# Structural diversity of molecular nitrogen on approach to polymeric states

Alexander F. Goncharov[1], Iskander G. Batyrev[2], Elena Bykova[3], Lukas Brüning,[4] Huawei Chen[1,5], Mohammad F. Mahmood [5], Andew Steele[1], Nico Giordano,[6] Timofey Fedotenko[6], Maxim Bykov[4,7]

[1] Earth and Planets Laboratory, Carnegie Institution for Science, Washington, DC 20015, USA
[2] U.S. Army Research Laboratory, FCDD-RLW-WA, Aberdeen Proving Ground, Maryland 21005, USA
[3] Goethe-Universität Frankfurt am Main, Facheinheit Mineralogie, 60438 Frankfurt am Main, Germany
[4] Institute of Inorganic Chemistry, University of Cologne, Greinstrasse 6, 50939 Cologne, Germany
[5] Department of Mathematics, Howard University, Washington D.C., 20059, USA
[6] Deutsches Elektronen-Synchrotron DESY, Notkestr. 85, 22607 Hamburg
[7] Institute of Inorganic and Analytical Chemistry, Johann Wolfgang Goethe Universität Frankfurt, Max-von-Laue-Straße 7, D-60438 Frankfurt am Main, Germany

**This pdf file contains Supplemental Figures S1-S7 and Tables S1-S3.**

**Corresponding author:**

**agoncharov@carnegiescience.edu**

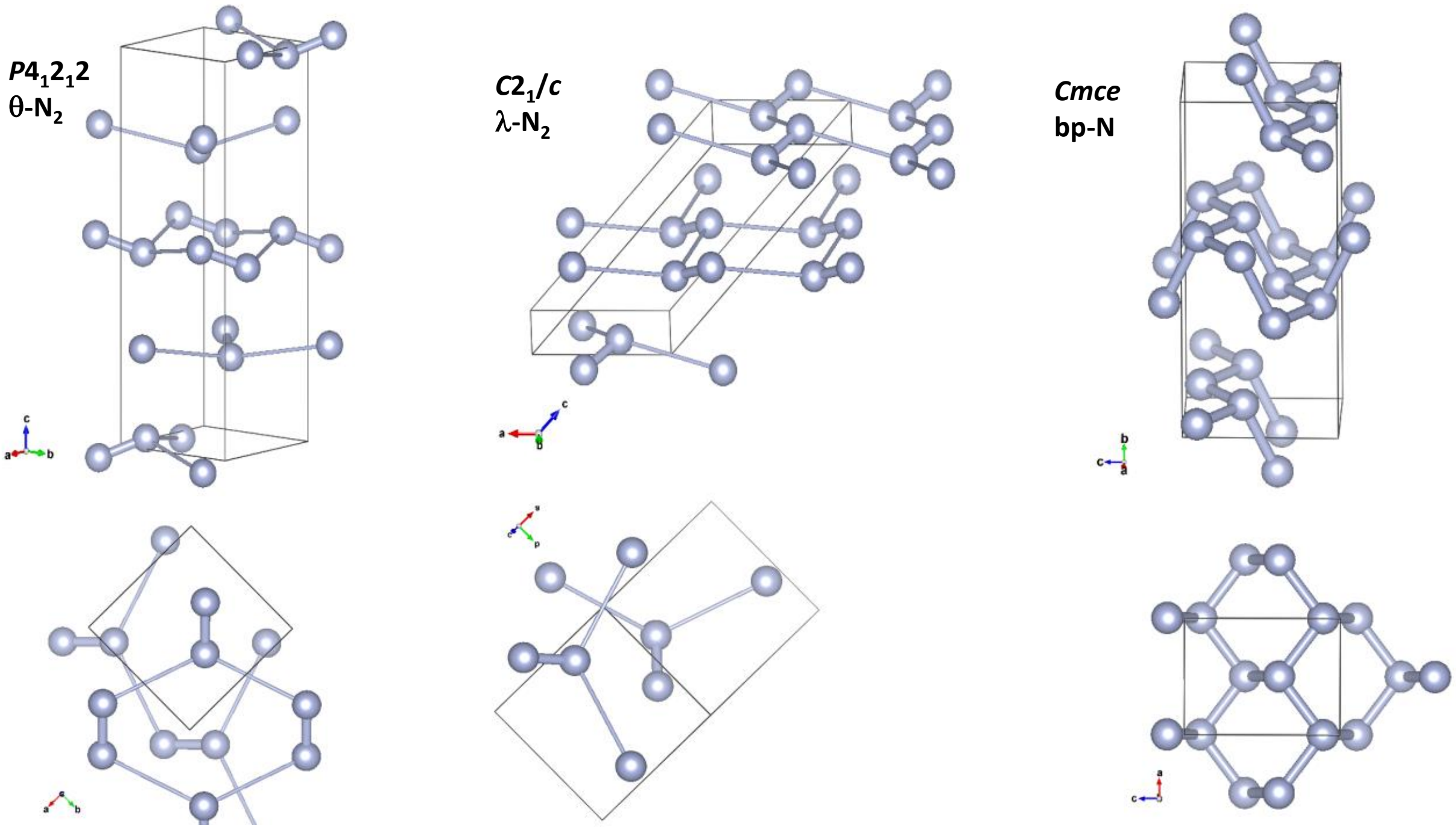

**Figure S1.** Crystal structures of nitrogen phases of interest to this work. Thick lines represent intramolecular N-N bonds and thin lines – intermolecular contacts. θ-$N_2$ and λ-$N_2$ are molecular phases, which consist of similar molecular layer but differ in their stacking. θ-$N_2$ consists of four layers forming an orthorhombic lattice. λ-$N_2$ consists of two layers stacked such that they form an orthorhombic lattice. BP-N is also layered, but all intralayer nearest neighbor bonds are covalent low order ones.

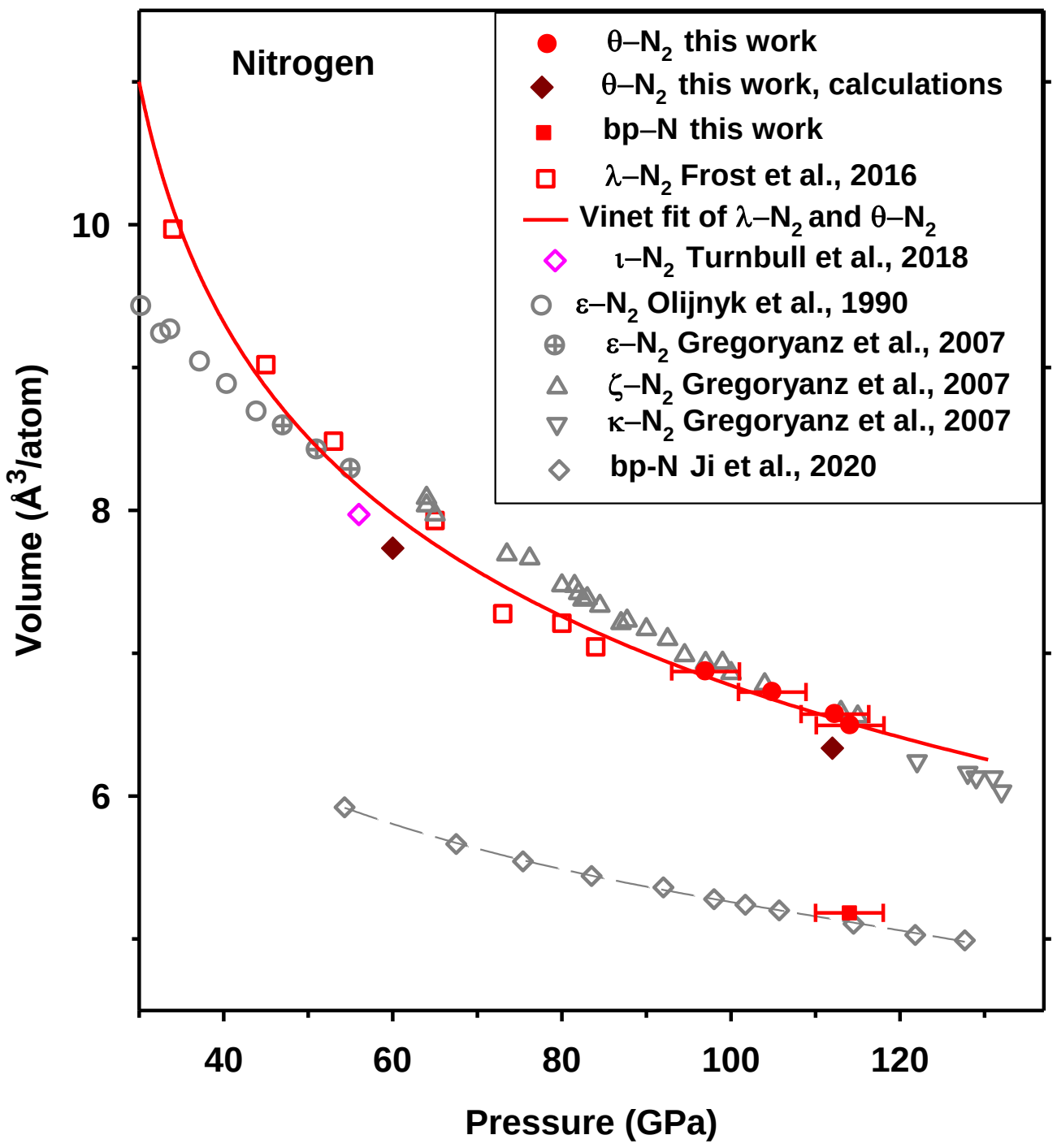

**Figure S2.** Volume-pressure data for different phases of nitrogen at 300 K. Filled symbols are the results of this work, open symbols are the results of previous experiments [9, 12, 22, 26, 39]. Solid line is a provisional isotherm in the form of the Vinet Equation of State (EOS) with $P_0$=30.0 GPa, $V_0$=11.0 Å$^3$, $K_0$=37(6) GPa and $K_0'$=6.0(7) drawn through the data for θ-$N_2$ of this work and λ-$N_2$ [26], which are expected to be alike due to the structural proximity (Fig. S1). At 114 GPa, our experiments show co-existence of three phases: molecular θ-$N_2$ and ζ-$N_2$ as well as polymeric bp-N.

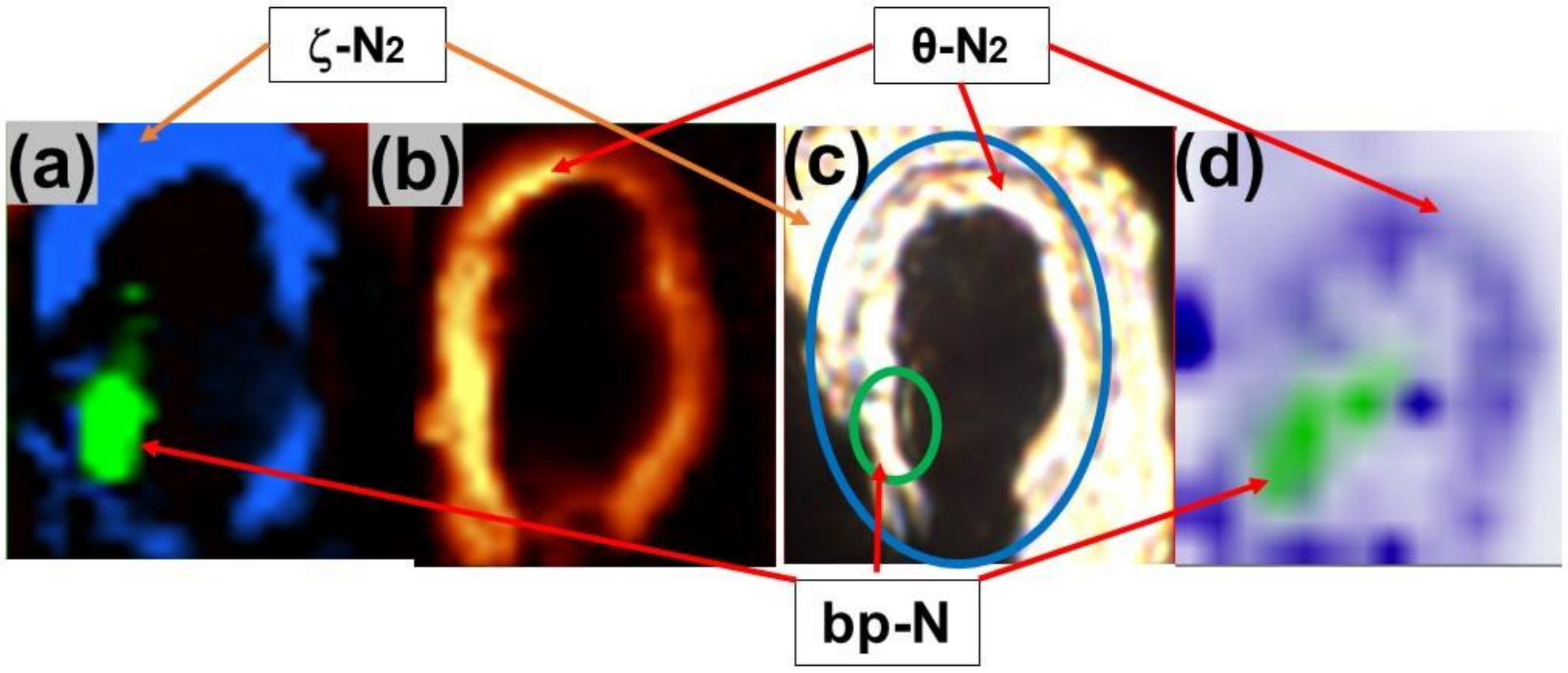

**Figure S3.** Images of the high-pressure cavity at 97 GPa after laser heating: (a) Raman mapping of bp-N and ζ-$N_2$; (b) Raman mapping of θ-$N_2$; (c) optical image; (d) X-ray mapping. The cavity is of approximately 50 μm in diameter and contains a piece of metallic Co metal (partially reacted to form a variety of nitrides) in the middle surrounded by nitrogen.

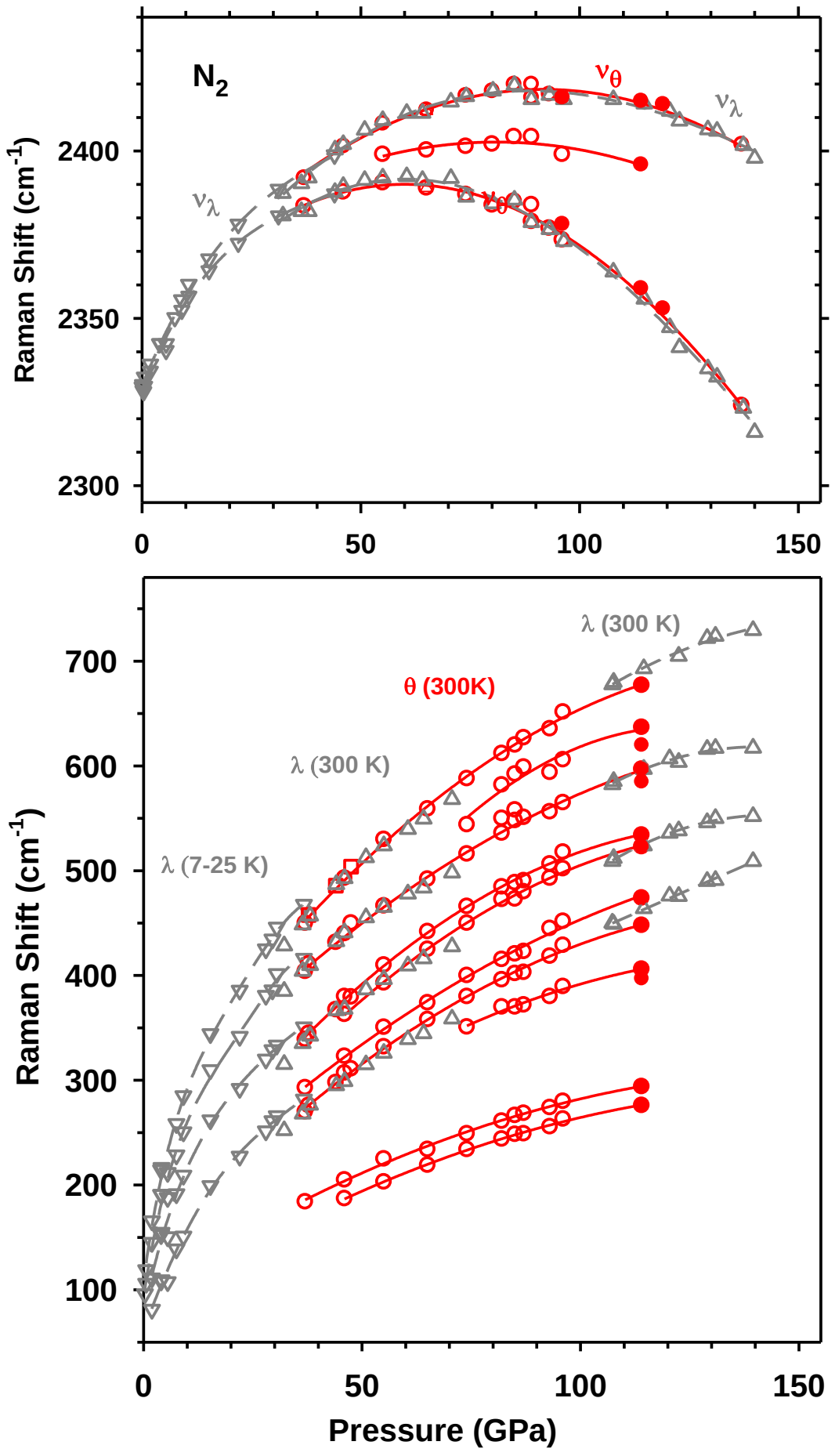

**Figure S4.** Raman frequencies vs pressure of θ-$N_2$ and λ-$N_2$. Red filled circles are the data of this work at 300 K; open symbols are the data of the previous works [23, 26]. Lines are guides to the eye.

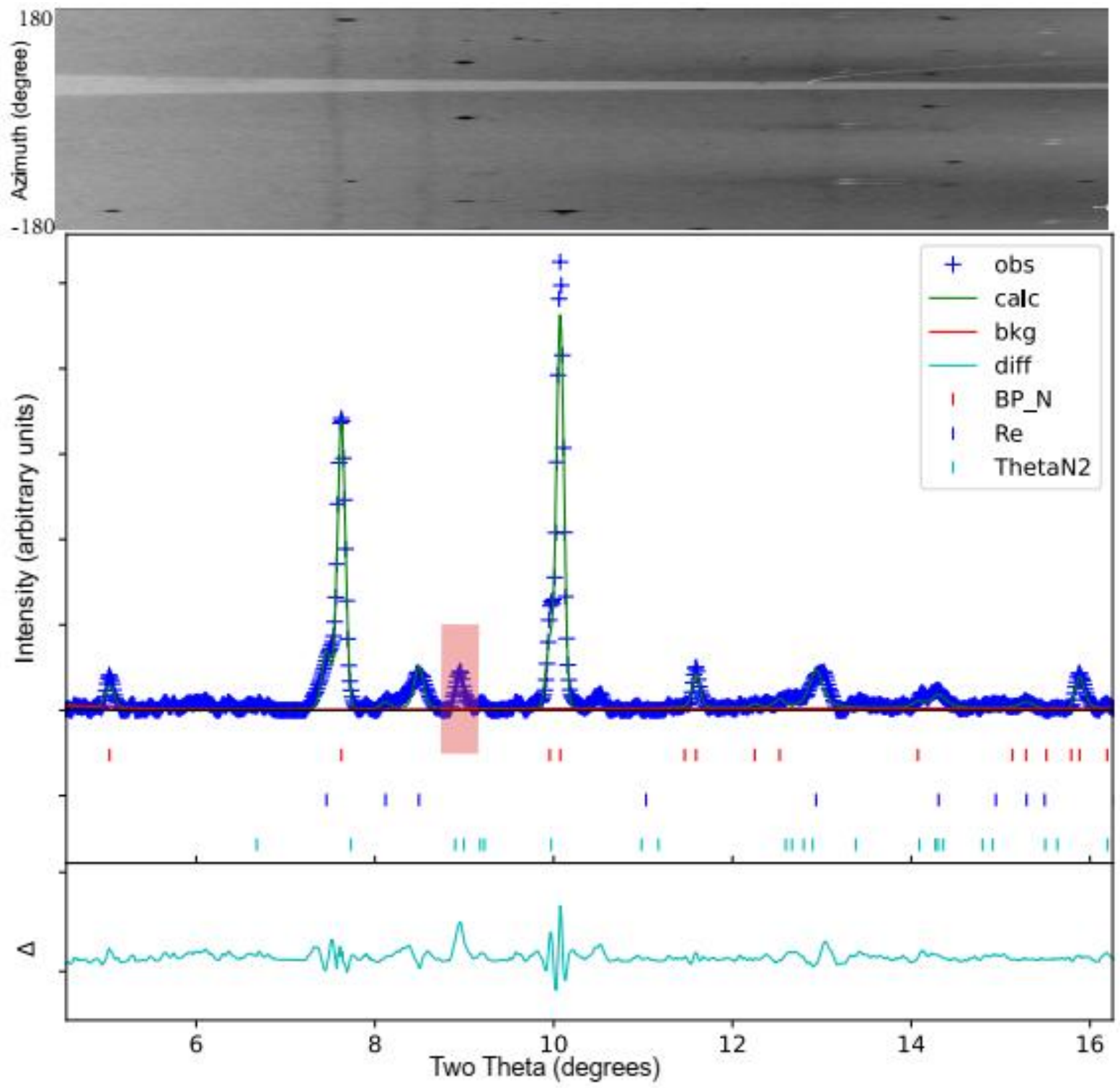

**Figure S5.** XRD pattern of bp-N crystallized after laser heating of θ-$N_2$ at 114 GPa. Crosses in the main panel are the experimental data (background subtracted), solid green line is a Le Bail fitting. Vertical ticks correspond the fitted positions of bp-N (*a*=2.156(3) Å, *b*=6.619(2) Å, *c*=2.905(6) Å, V=41.465(5) $Å^3$) and expected positions of other phases. The bottom curve is the difference curve between the experimental data and the fit. The top panel is the 2D diffractogram in rectangular coordinates (cake), which demonstrates SC nature of the Bragg peaks of bp-N. Red hatched area designates the position of the strongest peaks of θ-$N_2$. These peaks were not fitted because of the overall low-intensity of XRD peaks of θ-$N_2$.

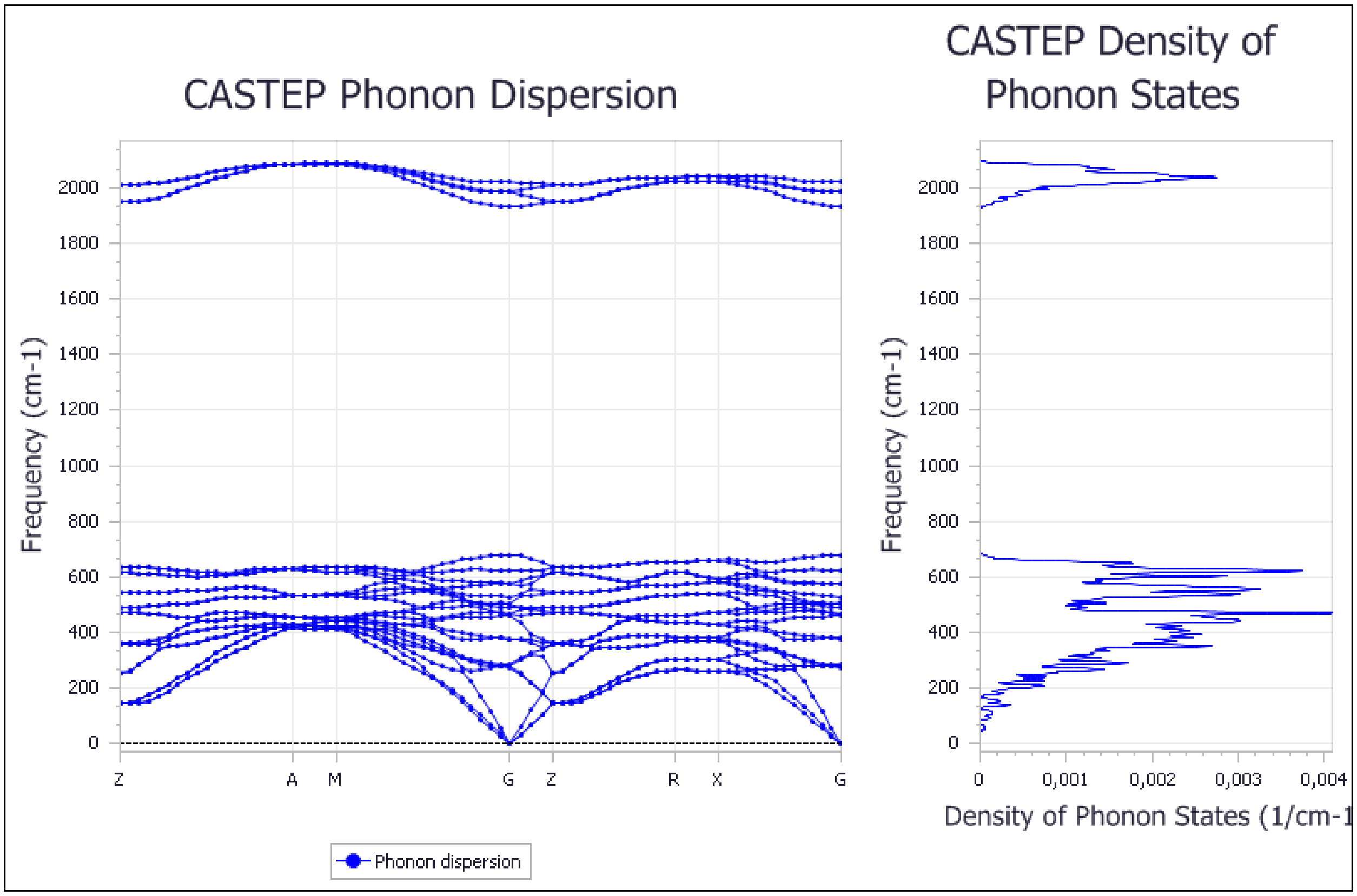

**Figure S6.** Theoretically calculated phonon dispersion curves of θ-$N_2$ at 112 GPa.

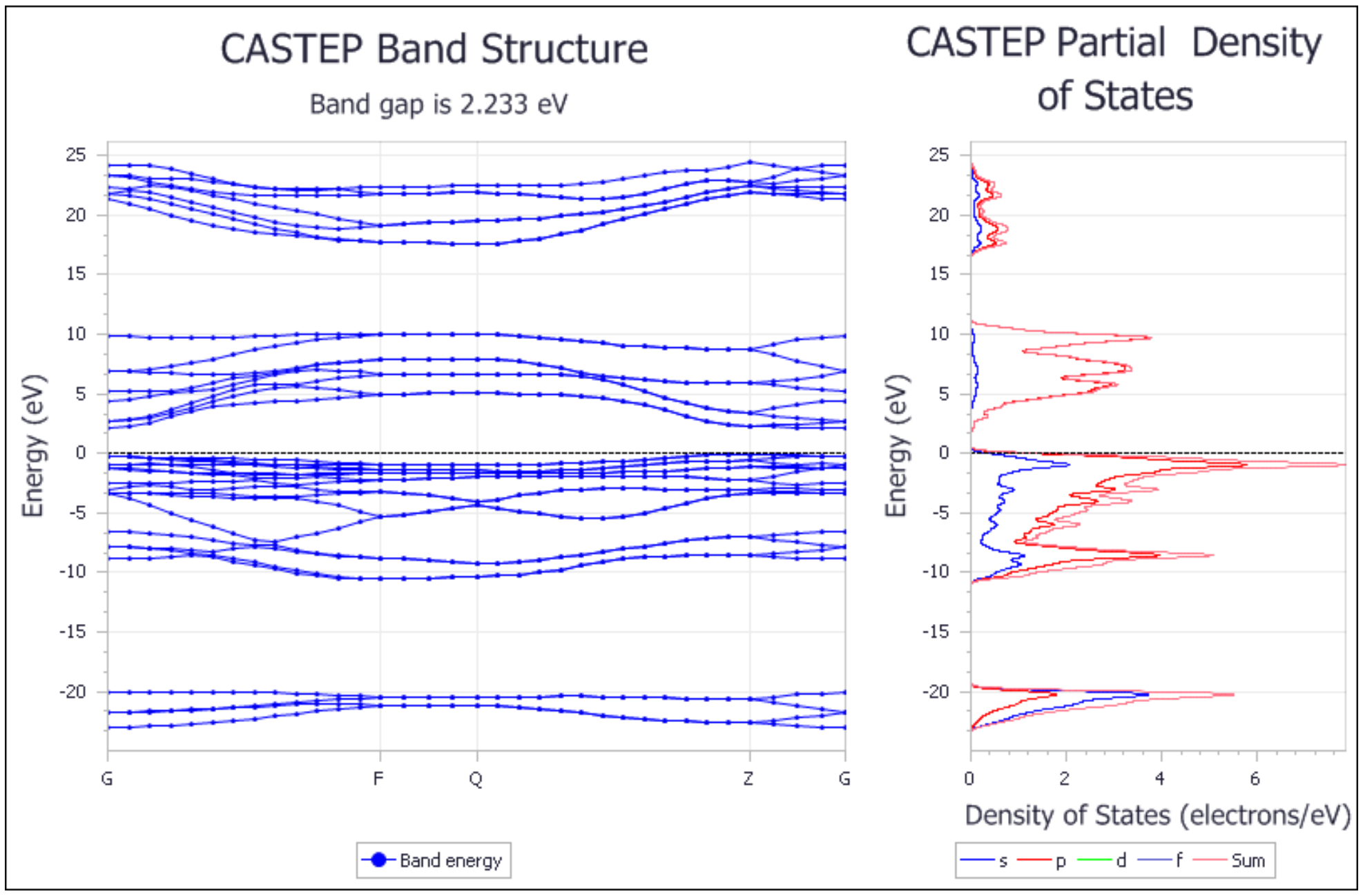

**Figure S7.** Theoretically calculated electronic band structure of θ-$N_2$ at 112 GPa.

## Table S1. Details of crystal structure refinements for θ-$N_2$ at high pressures

| Pressure (GPa) | 97 | 105 | 112 | 114 |
|---|---|---|---|---|
| Chemical formula | N | | | |
| $M_r$ | 14.01 | | | |
| Crystal system, space group | Tetragonal, $P4_12_12$ | | | |
| Temperature (K) | 293 | | | |
| *a*, *c* (Å) | 2.6943 (5), 7.52(3) | 2.6832(12), 7.48(9) | 2.6582(7), 7.44(1) | 2.6481(9), 7.41(3) |
| *V* (Å$^3$) | 54.6(3) | 53.9(6) | 52.59(9) | 52.0(2) |
| *Z* | 8 | | | |
| Radiation type | Synchrotron, λ = 0.29050 Å | | | |
| μ (mm$^{-1}$) | 0.08 | 0.08 | 0.08 | 0.08 |
| Diffractometer | LH @ P02.2 (Petra III, DESY) | | | |
| $T_{min}$, $T_{max}$ | 0.643, 1.000 | 0.504, 1.000 | 0.71764, 1.000 | 0.54297, 1.000 |
| No. of measured, independent and observed reflections | 207, 61, 47 [$I \geq 2\sigma(I)$] | 219, 62, 58 [$I \geq 2\ \sigma\ (I)$] | 224, 98, 90 [$I > 2\sigma(I)$] | 246, 81, 64 [$I > 2\sigma(I)$] |
| $R_{int}$ | 0.042 | 0.025 | 0.029 | 0.047 |
| $(\sin \theta/\lambda)_{max}$ (Å$^{-1}$) | 0.949 | 0.969 | 0.999 | 1.070 |
| **Refinement** | | | | |
| $R[F^2 > 2\sigma(F^2)]$, $wR(F^2)$, *S* | 0.042, 0.111, 1.25 | 0.065, 0.143, 1.10 | 0.050, 0.126, 1.16 | 0.040, 0.077, 1.15 |
| No. of reflections | 61 | 62 | 98 | 81 |
| No. of parameters | 10 | 10 | 10 | 10 |
| $\Delta\rho_{max}$, $\Delta\rho_{min}$ (e Å$^{-3}$) | 0.16, −0.16 | 0.49, −0.25 | 0.54, −0.30 | 0.32, −0.18 |

Computer programs: *CrysAlis PRO* 1.171.41.123a (Rigaku OD, 2022) [28] , SHELXT 2018/2 [29], olex2.refine 1.5-ac5-024 [40], *SHELXL2018*/3 [29], Olex2 1.5-ac5-024 [30].

**Table S2. Vibrational modes and their Raman and IR activity in $P4_12_12$ ($D_4^4$) and $P2_1/c$ ($C_{2h}^5$) crystal structures**

| Space & point group | θ-$N_2$ $P4_12_12$ #92 ($D_4^4$) | | λ-$N_2$ $P2_1/c$ #14 ($C_{2h}^5$) | |
|---|---|---|---|---|
| Site symmetry | 8b ($C_1$) | | 4e ($C_1$) | |
| Acoustic modes | $A_2$+E | | $A_u$+2$B_u$ | |
| Optical modes | Modes | Activity | Modes | Activity |
| | 3$A_1$+3$B_1$+3$B_2$+5E<br>2$A_2$+5E | Raman<br>IR | 3$A_g$+3$B_g$<br>3$A_u$+3$B_u$ | Raman<br>IR |
| Intramolecular | $A_1$+$B_1$+E | Raman+IR | $A_g$+$B_g$ | Raman |
| Translational | $A_1$+$B_1$+$B_2$+3E | Raman+IR | none | |
| Librational | $A_1$+$B_1$+2$A_2$+2$B_2$+E | Raman+IR | 2$A_g$+2$B_g$ | Raman |

**Table S3. Theoretically calculated Elastic Stiffness Constants $C_{ij}$ (GPa) of θ-$N_2$ at 112 GPa*)**

| | | | | | |
|---|---|---|---|---|---|
| 1059(5) | 666.2(1.3) | 605.8(1.7) | 0.00000 | 0.00000 | 0.00000 |
| 666.2(1.3) | 871(5) | 571.6(1.4) | 0.00000 | 0.00000 | 0.00000 |
| 605.8(1.7) | 571.6(1.4) | 926(4) | 0.00000 | 0.00000 | 0.00000 |
| 0.00000 | 0.00000 | 0.00000 | 168.3(7) | 0.00000 | 0.00000 |
| 0.00000 | 0.00000 | 0.00000 | 0.00000 | 296.8(1.1) | 0.00000 |
| 0.00000 | 0.00000 | 0.00000 | 0.00000 | 0.00000 | 294.7(1.4) |

*) Born conditions are satisfied: $C_{11} > C_{12}$; $2C_{13}^2 < C_{33}(C_{11} + C_{12})$; $C_{44} > 0$; $C_{66} > 0$